\documentstyle[pra,aps,amsfonts,]{revtex}

\begin{document}

\draft
\twocolumn[\hsize\textwidth\columnwidth\hsize\csname 
@twocolumnfalse\endcsname

\title{A comparative study of high-field diamagnetic fluctuations in deoxygenated $YBa_2Cu_3O_{7-x}$ and polycrystalline $(Bi-Pb)_2Sr_2Ca_2Cu_3O_{10}$}
\author{S. Salem-Sugui Jr.$^{1\ast}$, A. D. Alvarenga$^2$, K. C. Goretta$^3$, V. N. Vieira$^4$ ,B. Veal$^5$\\ and A. P. Paulikas$^6$}
\address{$^1$ Instituto de Fisica, Universidade Federal do Rio de Janeiro, 21945-970 Rio de Janeiro, Brasil\\
$^2$ Centro Brasileiro de Pesquisas Fisicas, Rua Dr. Xavier Sigaud 150, Rio de Janeiro, Brazil.\\
$^3$ Energy Technology Division, Argonne National Laboratory, Argonne, IL 60439, USA\\
$^4$ Centro de Ciencias Exatas e Tecnologia, Universidade de Caxias do Sul, RS, Brazil.\\
$^5$ Materials Science Division, Argonne National Laboratory, Argonne, IL 60439, USA}

\date{\today}

\maketitle
\begin{abstract}
We studied three single crystals of $YBa_2Cu_3O_{7-x}$ (Y123), with superconducting transition temperature, $T_c$= 62.5, 52, and 41 K,
and a highly textured polycrystalline specimen of $(Bi-Pb)_2Sr_2Ca_2Cu_3O_{10}$
(Bi2223), with $T_c$ = 108 K. Isofield
magnetization data were obtained as a function of temperature, with
the magnetic field applied parallel to the $c$ axis of each sample. The
reversible magnetization data for all samples exhibited a rounded transition as magnetization tended toward zero. The reversible data were interpreted in terms of two-dimensional diamagnetic lowest-Landau-level (LLL) fluctuation theory. The LLL scaling analysis yielded consistent values of the superconducting transition temperatures $T_{c}(H)$ for the various samples. The resulting scaling data were fit well by the two-dimensional LLL expression for magnetization obtained by Tesanovic and colaborators, producing reasonable values of $\kappa$ but the fitting parameter $\frac{\partial H_{c2}}{\partial T}$ produced values that were larger than the experimentally determined ones. We performed simultaneous scaling of Y123 data and Bi2223, obtaining a single collapsed curve. The single curve was obtained after multiplying the $x$ and $y$ axis of each scaling curve by appropriate sample-dependent scaling factors. An expression for the two-dimensional $x$-axis LLL scaling was extracted from theory, allowing comparison of theoretical values of the $x$-axis scaling factors with the experimental values.  The comparison between the values of the $x$-axis produced a deviation of 40$\%$ which suggests that the hypothesis of universality of the two-dimensional lowest-Landau-level fluctuations is not supported by the studied samples. We also observe that Y123 magnetization data for temperatures above $T_c$ obbey a universal scaling obtained for the diamagnetic fluctuation magnetization from a theory considering non-local field effects. The same scaling was not obbeyed by the corresponding magnetization calculated from the two-dimensional lowest-Landau-level theory.\\pacs{74.40.+k}
\end{abstract}
]

\section{introduction}
 
Fluctuation effects below and above the superconducting transition temperature
, $T_c(H)$, for high-$T_c$ superconductors have been the subject of
intensive work \cite {Lee,ullah,klemm,welp,tesa1,bula,tesa2,said1,said2,zacharias,moloni,rosenstein}. Field effects have been shown to enhance the fluctuation
region (beyond Gaussian) due to the one-dimensional character imposed by a
strong field to the Cooper pairs \cite {bergman,lee}, constraining the paired
quasi-particles to remain in the lowest Landau level (LLL). The
corresponding quartic term correction to the free energy density suppresses
the sharp second-order phase transition, rounding it on $\Delta T$ near $T_c(H)$ \cite{ikeda}. High- $T_c$ superconductors, with their small coherence lengths, $\xi $, high Ginzburg-Landau (GL) parameters,
$\kappa $, and high critical temperatures, $T_c$, display a broad
fluctuation region. Theoretical work has shown \cite {ullah,klemm} that in the
presence of strong magnetic fields, for which the LLL
approximation can be used, the $\Delta T$ fluctuation region around $T_{c}$, for
a system with dimensionality D is proportional to $(TH)^{(D-1)/D}$, and
that within this
region physical quantities such as magnetization and direct-current conductivity should be scaled with $(TH)^{(D-1)/D}$. For magnetization in
particular, scaling predicts that $M$ {\it vs} $T$ data obtained at
different fields, $H$, should collapse onto a single curve when the variable 
$M/(TH)^{(D-1)/D}$ is plotted against $(T-T_c(H))/(TH)^{(D-1)/D}$. Here, $T_{c}(H)$ becomes a fitting parameter. This scaling law has been used to
identify LLL fluctuations in a given material, and also to determine its
dimensionality \cite {welp,tesa1,bula,tesa2,said1,said2,moloni,rosenstein,said3}.\\
An important check of the
scaling is that it should provide reasonable values of $T_c(H)$ \cite {welp,said1,said3}. For lower-dimensional or layered materials, the LLL analysis also
helps to explain the crossing points observed in magnetization-vs-temperature (M-vs-T) curves \cite {tesa1,bula,said2,zacharias,rosenstein}.\\
Despite the intensive work on LLL fluctuations, there have been few experimental
efforts \cite {said3} devoted to study of the universality of the high-field fluctuation scaling, which is an important prediction of the LLL-based
theories for systems with a given dimensionality. Theories predicting a universal curve have been developed by Ullah and Dorsey \cite{ullah}, Tesanovic et al \cite{tesa1,tesa2}, and Rosenstein et al. \cite{rosenstein}. They have been used to fit data of specific systems. In most of the papers where a universal curve was fitted over experimental data, the fitting produced very large values of $\frac{\partial H_{c2}}{\partial T}$, values much larger (4-5 times) than experimentally determined ones.  Possibly due to this fact, the literature is to our knowledge lacking on experimental work that tests adequately the existence of a universal curve that should fit two or more different samples.  Furthermore, it is of fundamental interest to ask whether scaling for two different high-$T_c$ systems exhibits universal behavior. To provide such a test was the main objective of this work. The work was also motivated by the possibility of studying the evolution of the magnetic phase diagram of Y123 with
content of oxygen.\\ 
In this work, we addressed the above question by studying two different systems ($YBa_2Cu_3O_{7-x}$ (Y123) and a highly textured polycrystalline specimen of $(Bi-Pb)_2Sr_2Ca_2Cu_3O_{10}$
(Bi2223)).  These two systems are expected to exhibit D = 2 for high fields. We fit a two-dimensional LLL universal curve to the experimental data from each sample, which produced reasonable values of $\kappa$ for three samples (except, for the Y123 sample with $T_c$ = 62.5 K), but values of $\frac{\partial H_{c2}}{\partial T}$ were too large when compared with the experimentally determined values. We also performed a simultaneous two-dimensional LLL scaling analysis, obtaining a single collapsed curve for all data sets. The single collapsed curve was obtained after multiplying each sample data set by an appropriate sample-dependent scaling factor. An expression of the universal two-dimensional x-axis scaling factor was extracted from the theory and compared with the experimental values. The largest deviation found from this comparison is over 40$\%$. The values of $\frac{\partial H_{c2}}{\partial T}$ used in the theoretical x-axis expression were the experimental ones.  \\
\section{experiment}
The samples that were measured were: three single crystals of $YBa_2Cu_3O_{7-x}$
with critical temperatures, $T_c$ of 62.5 K (x = 0.35), 52 K (x = 0.5), and 41 K (x = 0.6), and corresponding masses of 1.2, 1.0, and 1.0 mg, and a highly textured polycrystalline $(Bi-Pb)_2Sr_2Ca_2Cu_3O_{10}$ with $T_{c}$ = 108 K \cite{goretta} and m $\simeq$ 4 mg. All of the Y123 single crystals were of approximate dimensions
1 x 1 x 0.2 mm, with the $c$ axis along the shorter direction, and each exhibited sharp, fully developed transitions ($\Delta T_c\simeq1K$). The
textured Bi2223 was cut into a similar shape as that of the single
crystals, but with a mass about four times larger. The Bi2223 sample also exhibited
a fully developed transition ($\Delta T_c\sim3K$).\\A commercial magnetometer (MPMS Quantum Design) based on a
superconducting quantum interference device (SQUID) 
was utilized. The scan length was 3 cm, which minimized field inhomogeneities. Experiments were conducted by
obtaining isofield magnetization (M) data as a function of temperature (T),
producing M-vs-T curves, with values of field running from 5000 to 50000 Oe.
Magnetization data were always taken after cooling the sample below $T_{c}$ in zero magnetic field. After cooling, a magnetic field was carefully applied (without
overshooting), always along the $c$ axis direction of the samples, and M-vs-T
curves were obtained by heating the sample to temperatures well above $T_c$, for fixed $\Delta T$ increments.\\We also obtained several field-cooled curves, corresponding to cooling the sample from above $T_c$ to below $T_c$, in an applied magnetic
field. This procedure allowed determination of the reversible (equilibrium)
magnetization. \\
\section{results and discussion}
Figure 1 shows zero-field-cooled M-vs-T data obtained for the Y123 (main figure) and Bi2223 (inset of Fig. 1) samples after background correction. Background corrections for all samples were of the type M(T) = A + B/T, where A and B are constants determined by fitting a selected region of a given curve well above $T_{c}(H)$. Two facts, observed for all curves of Y123 samples, are evident:\\ 

(1) Magnetization showed a "hump" that appeared at lower temperatures, occurring at the position at which the field-cooled and zero-field-cooled curves separated (where irreversibility set in). Such a hump appeared in all M(T) curves obtained from the Y123 samples. The hump is reminiscent of the so-called "paramagnetic Meissner effect",PME, observed in Type-II superconductors \cite{koshe,Pust}.  In this work, magnetization was obtained through a routine that fits a standard symmetric dipole into the output signal, which was generated after the entire length of the scan was completed. We observed this output dipole signal on the screen during some measurements. The dipole diamagnetic signal for the scan length that was used was symmetric for temperatures
corresponding to the reversible region of the M-vs-T curve, but the signal
dramatically lost symmetry for temperatures corresponding to the hump (irreversible region). Therefore, we concluded that the hump was probably an artifact of the equipment that was used and was due to the fitting routine. It is worth mentioning that the paramagnetic Meissner effect was also observed in niobium in Ref.19 where it is suggested that the PME is a surface effect not specific to high-$T_c$ superconductors. In this work, the hump was used to identify the lower-temperature limit of the reversible region in each M-vs-T curve.  We shall not discuss irreversibility effects or the irreversibility line in this work.\\ 
(2) A rounded transition was observed as magnetization tended to zero. Such a transition impeded a straightforward determination of $T_c(H)$. This response was observed in all M-vs-T curves that were obtained. Such a rounded transition has been shown to be associated with high-field
diamagnetic fluctuations of LLL type \cite {klemm,welp,tesa1,bula,said3}. Figure 1 also shows the existence of a crossing point in each data set, occurring close to $T_c(H)$ (clearest for the sample with $T_c$ = 41 K). These crossing points have been discussed previously for deoxygenated Y123 samples \cite{said4}. In that study \cite{said4}, the authors used measurements obtained for two samples of this work (the samples with $T_c$ = 62.5 K and 52 K).  \\ For the Y123 system, reduction of oxygen content decreases $T_c$ and increases the anisotropy $\gamma $ \cite{lifang}. For layered
systems, the crossing magnetic field above which the dimensionality of the system is reduced is expected to decrease as anisotropy increases \cite{klemm}. Because the value of the anisotropy of Bi2223 ($\gamma \sim 31-50 $ \cite{farrel,matsubara,clayton}), which is known to obey two-dimensional LLL scaling \cite {tesa1}, is of the same order of magnitude as that of $\gamma $ of our deoxygenated Y123 samples \cite{lifang}, we began analysis of the Y123 data by performing a LLL scaling with D = 2. We mention that we also attempted to perform a three-dimensional version of the
LLL scaling of our Y123 data.  The three-dimensional version was, however, shown to apply only to a narrow region close to $T_c(H)$ for the samples with $T_c$ = 52 and 62.5 K, and it is not presented here.\\
In the two-dimensional LLL analysis that was performed, $M/(TH)^{1/2}$
was plotted versus $(T-T_c(H))/(TH)^{1/2}$. $T_c(H)$ became a fitting parameter, chosen to make the data collapse onto
a single curve. Figure 2 shows the results of the scaling analysis performed on reversible data of Fig. 1. The values of the pairs $(T_{c}(H)(K),H (10^{3}Oe))$ and the correspondent value of $\frac{\partial H_{c2}}{\partial T}$ for each sample as obtained from the two-dimensional LLL analysis are listed in the table below: 

$$\left[\begin{array}{lccr}
Y123(41K)&Y123(52K)&Y123(62K)&Bi2223\\
(41,0)&(52,0)&(62.5,0)&(108.5,0)\\
(40.5,5)&(50.8,10)&(61.5,10)&(107,10)\\
(40,10)&(48.7,20)&(60.7,20)&(105.5,20)\\
(39.3,20)&(47.3,30)&(59.4,30)&(104,30)\\
(38.2,30)&(46,40)&(58.2,40)&(102.5,40)\\
(37,40)&(44.7,50)&(57,50)&(101,50)\\
(36,50)& & & \\
\frac{\partial H_{c2}}{\partial T}&\frac{\partial H_{c2}}{\partial T}&\frac{\partial H_{c2}}{\partial T}&\frac{\partial H_{c2}}{\partial T}\\
-10 kOe/K&-7 kOe/K&-9 kOe/K&-7.6 kOe/K
\end{array}\right]$$

It is worth mentioning that these values of $T_c(H)$ are in reasonable agreement with values of $T_c(H)$ estimated for each M-vs-T curve from the standard linear extrapolation of the reversible magnetization down to M = 0 \cite{abrikosov}. The obtained value of $\frac{\partial H_{c2}}{\partial T}$ for Bi2223 is in good agreement with published values \cite{matsubara,pekala}.\\ The results of the two-dimensional LLL analysis presented in Fig. 2 show that the collapse of
data was better for fields in the region of 20000-50000 Oe for the crystal with $T_c$ =
62.5 K, for fields in the region 10000-50000 Oe for the crystal with $T_c$ = 52 K, and for fields in the region 5000-50000 Oe for crystal with $T_c$ = 41 K. It is interesting to note that the lowest-field data obeying the two-dimensional LLL for Y123 decreased as the anisotropy of the system increased. This behavior is predicted by the theory. For layered
systems, the crossing magnetic field above which the dimensionality of the system is reduced is expected to decrease as anisotropy increases \cite{klemm}. The literature \cite{tesa1} presents an analytical expression for the two-dimensional LLL scaling function of the magnetization, which is given by Eq. 7 in Ref. 4 :
$$\frac{M(H,T)}{(HT)^{1/2}}\frac{s\phi_{0}}{A}|\frac{\partial H_{c2}}{\partial T}|=x-\sqrt{x^{2}+2} ,$$
where $x=A(T-T_{c}(H))/(TH)^{1/2}$, $A=a'(\phi_{0}s/2b)^{1/2}U_0$, $\phi_{0}$ is the quantum flux, $s$ is the interlayer distance, $U_0\sim0.8$ (around $H_{c2}(T)$), and $a'(T-T_c)=a(T)$, where $a$ and $b$ are the GL coefficients.\\ 
We applied this expression to our data. Fits by a least-squares minimization method were obtained with magnetization in Gauss units. Dashed lines in Fig. 2 represent the fitting results on the respective data. We used $d_{Y123}=6.38g/cm^3$ and $s_{Y123}=12\AA$. To our knowledge, the literature presents only one similar study (two-dimensional LLL fitting) performed in deoxygenated Y123, with $T_c\sim50 K$ \cite{rosenstein}. Bi2223 has been already studied and reported on in Ref.5. The fitting parameters are  $(\frac{\partial H_{c2}}{\partial T})_{T=Tc}$ and $a'/(2b)^{1/2}=\frac{1}{(\sqrt8\pi}(\frac{\partial H_{c}}{\partial T})_{T=Tc}=\frac{1}{4\sqrt\pi\kappa}(\frac{\partial H_{c2}}{\partial T})_{T=Tc}$, where we used the relation $H_{c2}=\sqrt2 \kappa H_c$ obtained in the GL linear approximation \cite{tinkham}, where $H_c$ is the GL critical field. The fitting results are shown in Fig. 2 and produced the values listed below: 
$$\left[\begin{array}{lccr}
Y123(41K)&Y123(52K)&Y123(62K)&Bi2223\\
\kappa=79&\kappa=73&\kappa=114&\kappa=96\\
|\frac{\partial H_{c2}}{\partial T}|&|\frac{\partial H_{c2}}{\partial T}|&|\frac{\partial H_{c2}}{\partial T}|&|\frac{\partial H_{c2}}{\partial T}|\\
21652 Oe/K & 19875 Oe/K & 3858 Oe/K & 23224 Oe/K
\end{array}\right]$$
It is important to note two interesting results of the fittings. (1) It is consistently suggested that as the system becomes more two dimensional (as $T_c$ of Y123 drops), fitting produces a value of the parameter $\kappa$ that is closer to the experimentally estimated value. This fact is, from our knowledge, a new result.  (2) Despite the good quality of the fittings, the experimental values of the superconducting diagmagnetic fluctuations above $T_{c}(H)$ are smaller than the theoretical ones.  These facts were observed for data of Y123 samples with $T_c$ = 52 and 41 K and for Bi2223. The possibility that the theory overestimates thermal fluctuations on the magnetization of high-$T_c$ superconductors, as suggested by the data analysis, has been discussed in Ref. 28. Although the fittings produced curves which reproduced quite well the experimental data (see the dashed lines in Fig. 2), with reasonable values of $\kappa$ (with the exception for Y123 with $T_c$= 62.5 K), the resulting values of $(\frac{\partial H_{c2}}{\partial T})_{T=T_c}$ are  2-3 times larger than the corresponding experimental values that we determined. Because of this discrepancy, we then used a different approach to interrogate the predicted universality of the two-dimensional LLL scaling.\\
We present in Fig. 3 a simultaneous scaling curve of all samples data sets, including reversible data obtained for the Bi2223 textured sample. The resulting collapsed curve was obtained after multiplying the collapsed curve obtained from the two-dimensional LLL analysis of each sample, as shown in Fig.2,  by an appropriate sample-dependent scaling factor.  Specifically, the $x$ axis and $y$ axis of Y123 and Bi2223 data were multiplied by the factors: x(0.84) and y(2.04) for Y123 with $T_c$ = 62.5 K, x(0.70) and y(1.61) for Y123 with $T_c$ = 52 K, x(0.67) and y(1.79) for Y123 with $T_c$ = 41 K, and x(1.18) and y(0.69) for Bi2223. The collapse of data from the four different samples, belonging to two different systems, was good and at a first interpretation, may provide support to the notion of universality. The dashed lines in Fig. 3 (which appear as a single fitting curve in Fig. 3 because they are too close to be resolved individually) represent the fittings already presented in Fig. 2.\\
We can take a step further, and analyze the collapse shown in Fig. 3 by comparing the scaling variables
of each axis. Comparison of the "magnetization-axis" ($y$ axis) is not directly possible
because of differences in sample geometry. For example, small
differences in thickness between samples, as was the case for these measurements, give rise to different demagnetization
factors. In any case, one may observe that the scaling factors of the $y$ axis were roughly proportional to sample mass ratio. Comparison of the scaling variable along the $x$ axis is possible,
because Ullah and Dorsey \cite{ullah} provide an analytical expression for the full two-dimensional LLL temperature scaling variable along the $x$ axis: 
$$x=\frac{s}{(2\kappa^2-1)^{1/2}}|\frac{\partial H_{c2}}{\partial T}|\frac{(T-T_c(H))}{(TH)^{1/2}} ,$$ 
where s is the interlayer spacing and $\kappa$ is
the Ginzburg-Landau parameter. The ratio of the respective Bi2223 and Y123 scaling factors, 
$$r=(\frac{s}{(2\kappa ^{2}-1)^{1/2}}|\frac{\partial H_{c2}}{\partial T}|)_{Bi2223}/(\frac{s}{(2\kappa ^2-1)^{1/2}}|\frac{\partial H_{c2}}{\partial T}|)_{Y123}$$ 
can be used to rescale the Bi2223 data to make them collapse onto the Y123 data. The experimental values of $r=x_{Bi}/x_{Y123}$, where $x_{Bi}$ and $x_{Y123}$ are the $x$ axis sample-dependent
factors listed above. A comparison of both $r$ values for the samples that were examined can test the two-dimensional LLL hypothesis of universality.\\ 
For a direct comparison with the experimental values, we require values of $\kappa$, which for Y123 can be estimated from isothermal $MvsH$ curves previously obtained for our Y123 samples \cite{valdemar,peak}. The inset of Fig. 3 shows selected isothermal $M-vs-H$ curves. Measurements of additional $M-vs-H$ curves in these samples could not be made reliably because of the current low-field restrictions of our equipment. The curves in the inset of Fig. 3 allow one to estimate values of $H_{c1}$. This value was obtained as the field for which the magnetization curve deviates from the linear Meissner region. These values, coupled with the GL relation, $H_{c1}/H_{c2}=\ln{\kappa}/2\kappa^2$ and values of $H_{c2}$ obtained in this work from the 2D-LLL scaling analysis (values are listed above), then allow one to estimate values of $\kappa$. 
For the sample with $T_c$ = 62.5 K, we also estimate values of $H_{c1}$ from $MvsH$ curves obtained at low temperatures \cite{peak}, which allow us to obtain the extrapolated value of $H_{c1}(T=0)$. The estimated values of $\kappa$ are: $\kappa \sim65$ for Y123($T_c$=52 K), $\kappa \sim56$ for Y123($T_c$=62.5 K) and $\kappa \sim80$ for Y123($T_c$ = 41 K). The error on $\kappa$ is (over) estimated to be $\prec10$.\\ 
By using the experimental obtained values of 
$\frac{\partial H_{c2}}{\partial T}$, $\kappa$ for Y123, $s_{Y123}$=$12\AA$, $\kappa_{Bi}$ = 82 \cite{pekala}, and $s_{Bi}$=$20\AA$ in the expression above for $r$, we obtain: $r(\frac{Bi}{Y62K})$= 0.96 ,  $r(\frac{Bi}{Y52K})$= 1.43 and $r(\frac{Bi}{Y41K})$ = 1.23. The experimental values are $r(\frac{Bi}{Y62K})$ = 1.40, $r(\frac{Bi}{Y52K})$ = 1.69, and $r(\frac{Bi}{Y41K})$ = 1.76. The largest deviations between experimental and calculated values are $\sim 40\%$, which is found for samples with $T_c$= 62.5 and 41 K.\\ 
To provide conclusive evidence of two-dimensional LLL scaling universality, one would expect much smaller deviations in $r$. It should be noted that, independently of the above result, the basic scaling theory is expected to be independently valid for each system here studied. We finally note that in previous work with Nb \cite{said4}, where simultaneous scaling was performed for Nb and Y123 data (two systems with dimension = 3), evidence for universal three-dimensional LLL behavior was not obtained.\\
Finaly, we attempt to verify if diamagnetic fluctuations due to non-local electrodynamics effects \cite{payne}, that appear from strong magnetic fields, might explain the observed deviations between theory and experiment concerning two-dimensional LLL. It is important to mention that the LLL theory was developed under the Ginzburg-Landau theory for strong magnetic fields \cite{lee} which is local \cite{lee} and corresponds to the first term expansion of the Gorkov theory. For stronger fields \cite{payne}, higher order terms expansion of the Gorkov theory are necessary producing non-local field terms. In this case, it has been shown \cite{payne} that $M(T)H^{-0.5}/T$ is a universal function of $H/H_s$, where $M(T)$ is the fluctuation magnetization and $H_s$ is a sample dependent field. Experimentally, the above scaling only works at $T=T_c$. Above $T_c$, $M(T)H^{-0.5}/T$ vs $H/H_s$ produced universal curves at temperatures given by $\mid dH_{c2}dT\mid(T-T_{c})/H=c$ where $c\succ0$ is a constant. This universal behavior have been shown to work for many convencional type II superconductors \cite{payne,tinkham2,tinkham} with $H_s$ as high as $H_{c2}$, but, from our knowledge, it was never tested on high field data of high-Tc superconductors. Figure 4a shows curves $M(T)H^{-0.5}/T$ vs $H/H_s$ obtained for Y123 data shown in Fig. 1. We only analysed Y123 data because they have a sharper $T_c$ when compared to Bi2223 (in the analysis of convencional superconductor data, $\Delta T_{c}\approx10^{-3}K$ \cite{tinkham2}). It is possible to see in Fig. 4a that the behavior of the curve obtained with values of magnetization at $T=T_c$ is followed by the curves with $\mid dH_{c2}dT\mid(T-T_{c})/H=0,5$ and $1$, which resembles the behavior observed in the similar plot for convencional superconductors \cite{tinkham2}. It is not possible to verify if our data follow the same universal curve of Ref.32 since the magnitude of $M(T)H^{-0.5}/T$ for our samples are two orders larger than the same quantity for conventional superconductors. In plotting Fig. 4a, we try to obtain values of the X axis with similar order of magnitude (aproximately one order larger here) as in the universal curve obtained for convencional superconductors appearing in Ref.32, which produced the values of the field $H_s$ listed in Fig. 4a. It is interesting to note that the data for Y123 with $T_{c}=62 K$ do not follow the "scale" for the curves with $\mid dH_{c2}dT\mid(T-T_{c})/H=0,5$ and $1$. This is because data for this sample ($T_{c}=62 K$) have a (almost) perfect match with the 2D-LLL theory (see Fig. 2). At $T=T_c$, we observed that the experimental values of magnetization are close to the values calculated from the 2D-LLL expression, which means that only at $T=T_c$ the fluctuation magnetization obtained from the 2D-LLL theory also satisfies the non-local field scaling of Fig. 4a. The later fact does not occurr for temperatures above $T_c$. Figure 4b shows a plot for samples with $T_c$=52 and 41 K (similar to Fig. 4a) for $\mid dH_{c2}dT\mid(T-T_{c})/H=1$ , but with magnetization calculated from the 2D-LLL expression using the fitting results ($\mid dH_{c2}dT\mid$ and $\kappa$) listed in the table (the curves calculated for $\mid dH_{c2}dT\mid(T-T_{c})/H=0.5$ are very similar and are not shown). It is clear in Fig. 4b that, for temperatures above $T_c$, fluctuation magnetization calculated from 2D-LLL for samples with $T_c$=52 and 41 K do not follow the non-local field scaling that experimental data do in Fig. 4a. Comparison of Figs. 4a and 4b suggests that non-local electrodynamics effects due to high fields should have an important rule on fluctuation diamagnetism above $T_c$ for these two samples, which might explain the deviations between the experiment and lowest-Landau-level theory found in this work.\\ 
In conclusion, our study shown that high field fluctuations on reversible data of Y123 crystals and  Bi2223 are well explained by two-dimensional lowest-Landau-level fluctuations. The scaling analysis performed on Y123 data produced consistent values for the upper critical field $H_{c2}$ for the studied samples. The data of each sample was fitted by the predicted 2D-LLL universal theory, which produced reasonable values of the Ginzburg-Landau parameter $\kappa$, but much larger values of $\frac{\partial H_{c2}}{\partial T}$. The fitting show that theory predicts larger values for the diamagnetic fluctuation for temperatures above $T_c(H)$. A different test, than the fitting, was performed to check the hypothesis of universality of 2D-LLL, but the results did not support this hypothesis for the studied samples. We also observe that Y123 magnetization data for temperatures above $T_c$ obbey a universal scaling obtained for the diamagnetic fluctuation magnetization from a theory considering non-local field effects. The same scaling was not obbeyed by the corresponding magnetization calculated from the two-dimensional lowest-Landau-level theory.\\ 
We grateful thank the Referee who suggested us to verify the possible importance of non-local eletrodynamics effects on the fluctuation diamagnetism in our data. We also thank Mark Friesen for helpful discussions. This work was partially supported by CNPq, Brazilian Agency. The work at Argonne National Laboratory was supported by the U.S. Department of Energy, under Contract W-31-109-Eng-38.\\
$\ast$ Corresponding author. E-mail: said@if.ufrj.br

Figure Captions\\
Fig1. Zero-field-cooled M-vs-T curves as obtained for $YBa_2Cu_3O_{7-x}$ crystals for magnetic fields ranging from 5000 to 50000 Oe for crystal with $T_c$ = 41 K, and from 10000 to 50000 Oe for crystals with $T_c$ = 52 K and 62.5 K. The inset shows  zero-field-cooled magnetization data of Bi2223 used in the scaling analysis, obtained for magnetic fields ranging from 10000 do 50000 Oe.\\

Fig2. Two-dimensional LLL scaling of Y123 and Bi2223 (inset) data presented in Fig. 1. Dashed lines represent fittings of the data to the two-dimensional LLL universal expression for magnetization. Y123 curves were displaced along the Y axis to appear in the same figure.\\

Fig3. Simultaneous two-dimensional LLL scaling of Y123 and Bi2223 (inset) data presented in Fig. 1. Dashed lines represent fittings of the data to the two-dimensional LLL universal expression for magnetization. The inset shows selected isothermic zero-field-cooled magnetization curves for the Y123 crystals. The magnetization curves were displaced along the Y axis to appear in the same figure, and magnetization for Y123 with $T_c$= 41 K was multiplied by 2.\\

Fig.4 (a)$M(T)H^{-0.5}/T$ is plotted against $H/H_s$ for Y123 data of Fig. 1. Dashed lines are only a guide to the eyes. (b)$M(T)H^{-0.5}/T$ is plotted against $H/H_s$ with values of M(T) obtained from the two-dimensional lowest-Landau-level expression.

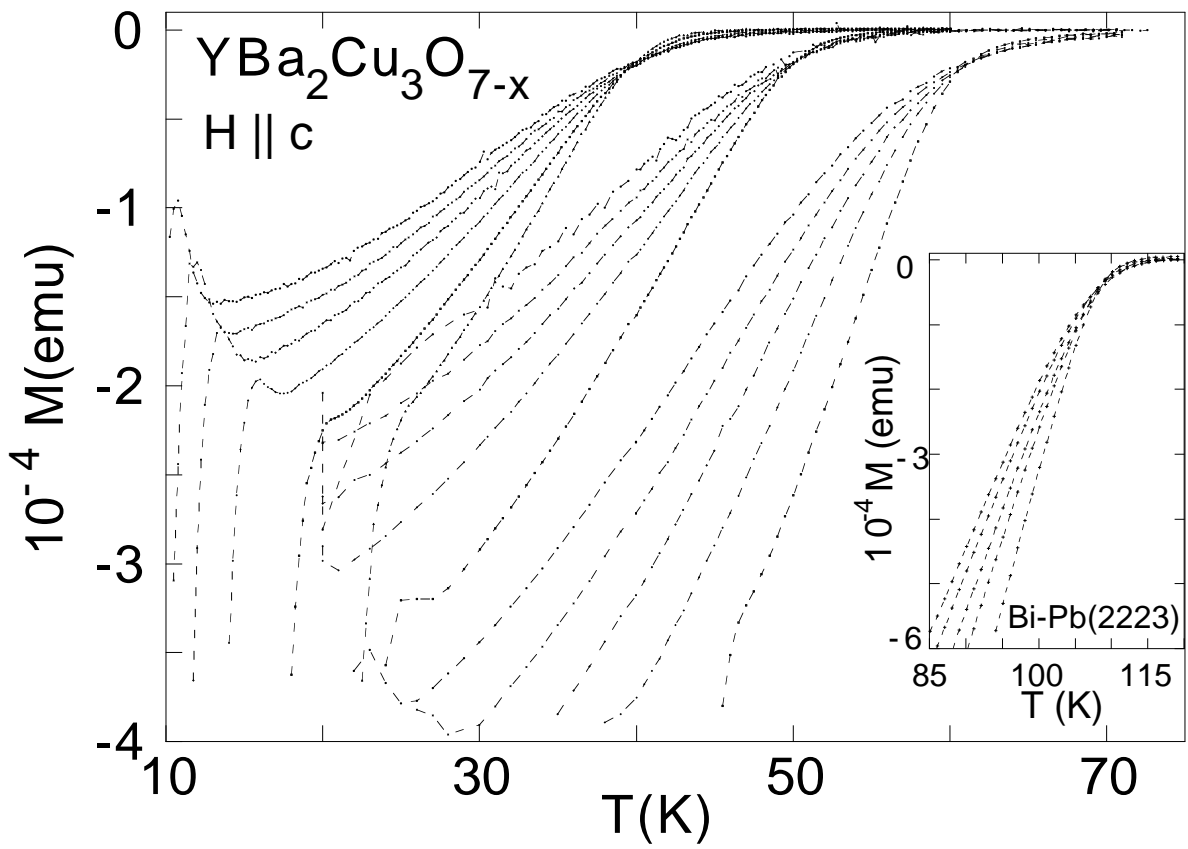
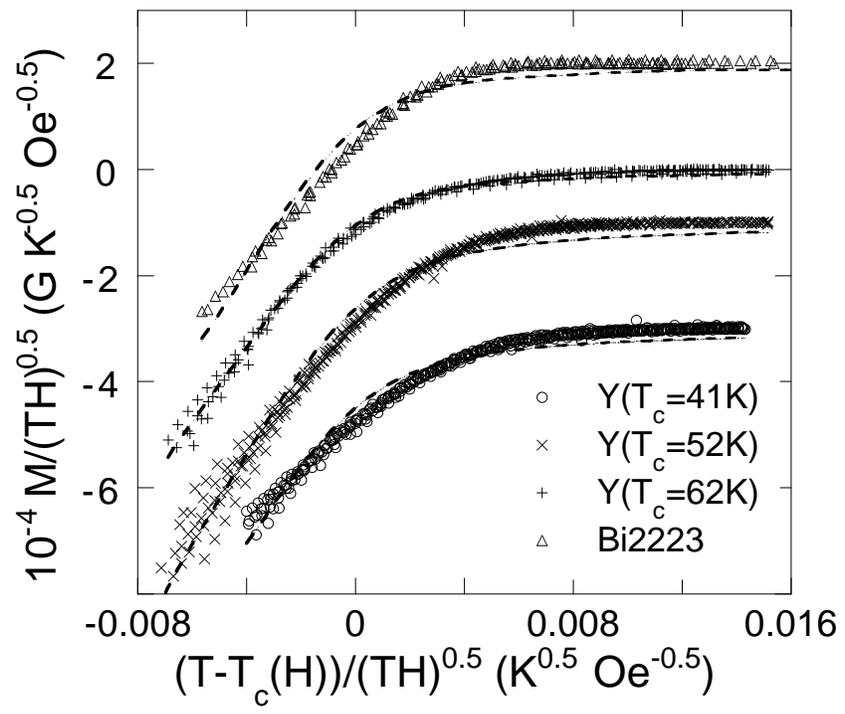
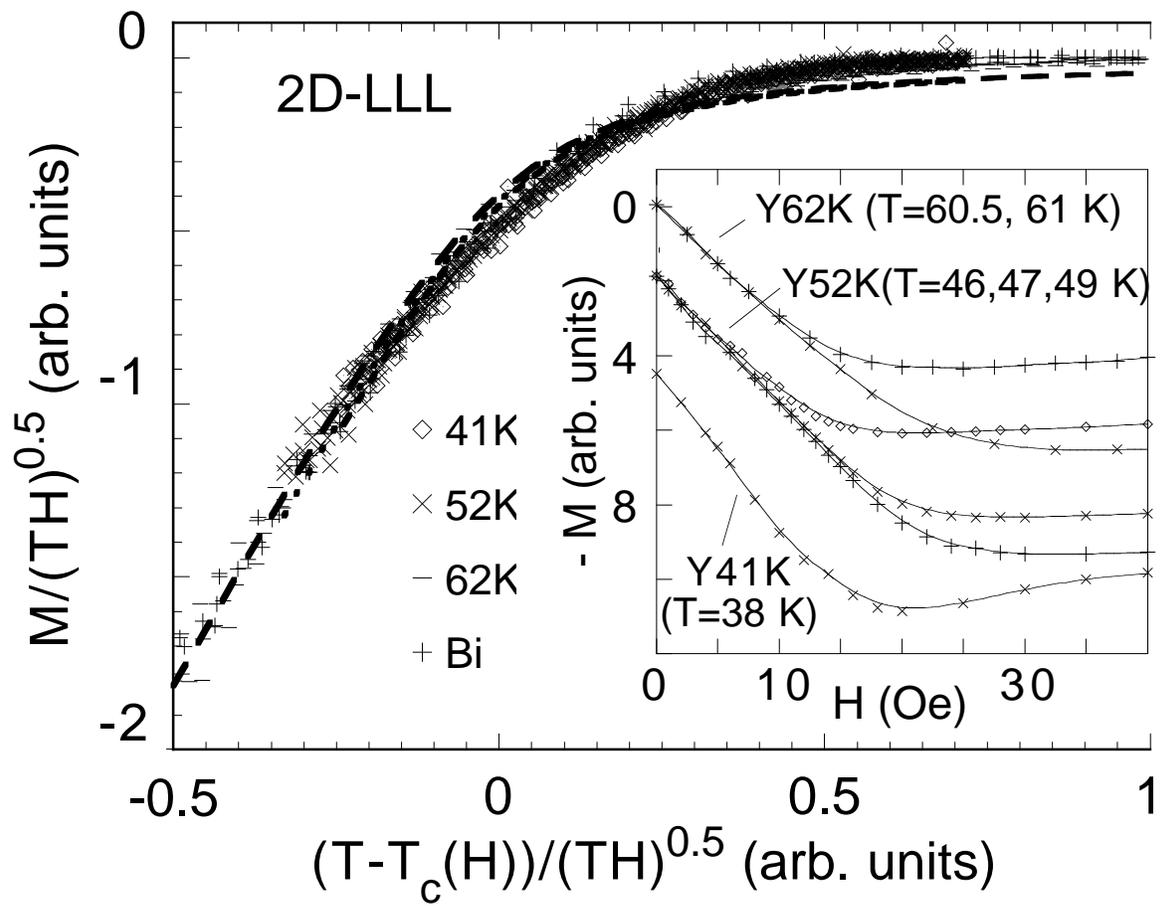
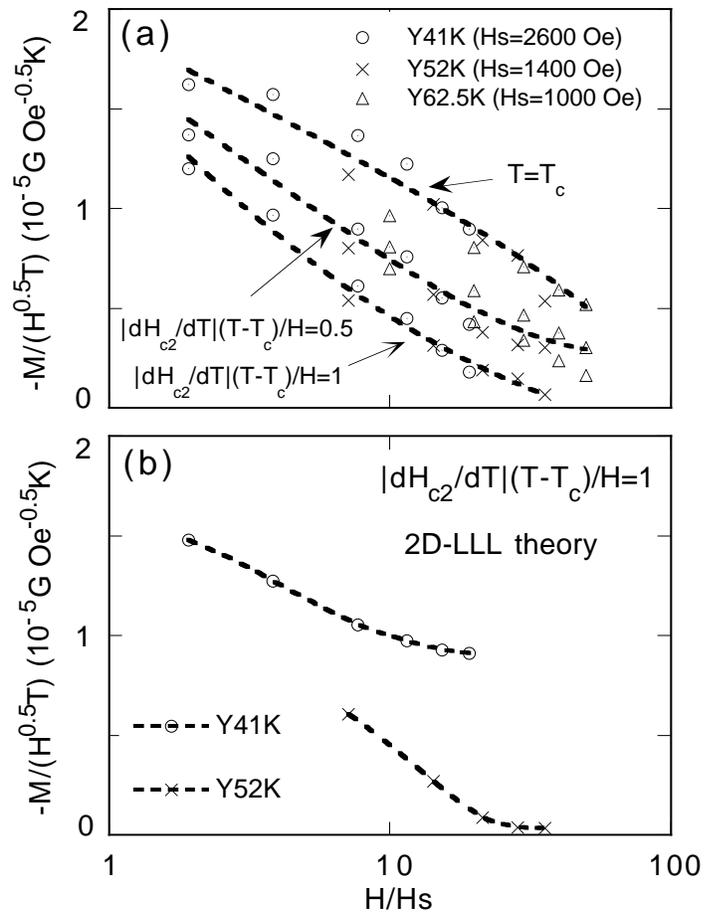
\end{document}